\documentclass[a4paper]{article}
\makeatletter\if@twocolumn\PassOptionsToPackage{switch}{lineno}\else\fi\makeatother

\usepackage{tabulary,graphicx}
\usepackage{float} 
\usepackage{gensymb} 

\usepackage[T1]{fontenc}
\usepackage[utf8]{inputenc}
\usepackage{csquotes} 
\usepackage[colorlinks,bookmarksopen=true,bookmarksnumbered,citecolor=black,urlcolor=black, linkcolor= black]{hyperref}

\usepackage{doi}
\usepackage{orcidlink}

\usepackage{natbib}

\usepackage{nameref}
\usepackage{bookmark}

\usepackage{url,multirow,morefloats,floatflt,cancel}
\makeatletter

\AtBeginDocument{\@ifpackageloaded{textcomp}{}{\usepackage{textcomp}}}
\makeatother
\usepackage{colortbl}
\usepackage{multirow}
\usepackage{makecell}
\usepackage{xcolor}
\usepackage{soul}
\definecolor{lightgray}{gray}{0.30}
\sethlcolor{lightgray}
\makeatletter
 \def\SOUL@hlpreamble{%
 \setul{}{3ex}
 \let\SOUL@stcolor\SOUL@hlcolor
 \SOUL@stpreamble
 }
\makeatother
\definecolor{c_statq}{HTML}{bfef45} 
\definecolor{c_statn}{HTML}{808000} 
\definecolor{c_autoq}{HTML}{ffe119} 
\definecolor{c_auton}{HTML}{f58231} 
\definecolor{c_transq}{HTML}{e6194b} 
\definecolor{c_transn}{HTML}{800000} 
\definecolor{c_videoq}{HTML}{F032e6} 
\definecolor{c_videon}{HTML}{911Eb4}
\def\gr{\textcolor{gray}}

\usepackage{pifont}
\usepackage[nointegrals]{wasysym}
\makeatletter

\def\mcWidth#1{\csname TY@F#1\endcsname+\tabcolsep}

\def\cAlignHack{\rightskip\@flushglue\leftskip\@flushglue\parindent\z@\parfillskip\z@skip}
\def\rAlignHack{\rightskip\z@skip\leftskip\@flushglue \parindent\z@\parfillskip\z@skip}

\@ifundefined{etal}{}{}

\usepackage{ifxetex}
\ifxetex\else\if@twocolumn\@ifpackageloaded{stfloats}{}{\usepackage{dblfloatfix}}\fi\fi

\AtBeginDocument{
\expandafter\ifx\csname eqalign\endcsname\relax
\def\eqalign#1{\null\vcenter{\def\\{\cr}\openup\jot\m@th
  \ialign{\strut$\displaystyle{##}$\hfil&$\displaystyle{{}##}$\hfil
      \crcr#1\crcr}}\,}
\fi
}

\AtBeginDocument{%
  \@ifpackageloaded{endfloat}%
   {\renewcommand\efloat@iwrite[1]{\immediate\expandafter\protected@write\csname efloat@post#1\endcsname{}}}{\newif\ifefloat@tables}%
}%

\def\BreakURLText#1{\@tfor\brk@tempa:=#1\do{\brk@tempa\hskip0pt}}
\let\lt=<
\let\gt=>
\def\processVert{\ifmmode|\else\textbar\fi}

\@ifundefined{subparagraph}{
\def\subparagraph{\@startsection{paragraph}{5}{2\parindent}{0ex plus 0.1ex minus 0.1ex}%
{0ex}{\normalfont\small\itshape}}%
}{}

\newcommand\role[1]{\unskip}
\newcommand\aucollab[1]{\unskip}
  
\@ifundefined{tsGraphicsScaleX}{\gdef\tsGraphicsScaleX{1}}{}
\@ifundefined{tsGraphicsScaleY}{\gdef\tsGraphicsScaleY{.9}}{}
\def\checkGraphicsWidth{\ifdim\Gin@nat@width>\linewidth
	\tsGraphicsScaleX\linewidth\else\Gin@nat@width\fi}

\def\checkGraphicsHeight{\ifdim\Gin@nat@height>.9\textheight
	\tsGraphicsScaleY\textheight\else\Gin@nat@height\fi}

\def\fixFloatSize#1{}
\let\ts@includegraphics\includegraphics

\def\inlinegraphic[#1]#2{{\edef\@tempa{#1}\edef\baseline@shift{\ifx\@tempa\@empty0\else#1\fi}\edef\tempZ{\the\numexpr(\numexpr(\baseline@shift*\f@size/100))}\protect\raisebox{\tempZ pt}{\ts@includegraphics{#2}}}}

\AtBeginDocument{\def\includegraphics{\@ifnextchar[{\ts@includegraphics}{\ts@includegraphics[width=\checkGraphicsWidth,height=\checkGraphicsHeight,keepaspectratio]}}}

\DeclareMathAlphabet{\mathpzc}{OT1}{pzc}{m}{it}

\def\URL#1#2{\@ifundefined{href}{#2}{\href{#1}{#2}}}

\def\UrlOrds{\do\*\do\-\do\~\do\'\do\"\do\-}%
\g@addto@macro{\UrlBreaks}{\UrlOrds}

\edef\fntEncoding{\f@encoding}

\makeatother

\newif\ifmultipleabstract\multipleabstractfalse%
\makeatletter\AtBeginDocument{\@ifpackageloaded{endfloat}{\usepackage{lastpage}}{}}\makeatother

\usepackage{float}

\begin{document}

\title{Effect of Avatar Head Movement on Communication Behaviour, Experience of Presence and Conversation Success in Triadic Conversations}


\author{
Angelika Kothe$^1$\orcidlink{https://orcid.org/0009-0009-0176-0881}, 
Volker Hohmann$^1$\orcidlink{https://orcid.org/0000-0001-7056-1880}, 
Giso Grimm$^1$\orcidlink{https://orcid.org/0000-0002-9685-4507}}

\maketitle

{$^1$ Department of Medical Physics and Acoustics, Carl von Ossietzky Universit\"at Oldenburg, and Cluster of Excellence ``Hearing4All'', Oldenburg, Germany}



\begin{abstract}
  Interactive communication in virtual reality can be used in experimental paradigms to increase the ecological validity of hearing device evaluations. This requires the virtual environment to elicit natural communication behaviour in listeners. This study evaluates the effect of virtual animated characters' head movements on participants' communication behaviour and experience. 
  Triadic conversations were conducted between a test participant and two confederates. To facilitate the manipulation of head movements, the conversation was conducted in telepresence using a system that transmitted audio, head movement data and video with low delay. The confederates were represented by virtual animated characters (avatars) with different levels of animation: Static heads, automated head movement animations based on speech level onsets, and animated head movements based on the transmitted head movements of the interlocutors. A condition was also included in which the videos of the interlocutors' heads were embedded in the visual scene. 
  The results show significant effects of animation level on the participants' speech and head movement behaviour as recorded by physical sensors, as well as on the subjective sense of presence and the success of the conversation. The largest effects were found for the range of head orientation during speech and the perceived realism of avatars. Participants reported that they were spoken to in a more helpful way when the avatars showed head movements transmitted from the interlocutors than when the avatars' heads were static. 
  We therefore conclude that the representation of interlocutors must include sufficiently realistic head movements in order to elicit natural communication behaviour.
\end{abstract}

keywords: communication behaviour, conversation success, head movement, audiovisual virtual reality, hearing device evaluation

\section*{Introduction} 
Conversations, possibly with surrounding background noise, occur frequently in every-day life \citep{wolters_common_2016}. They are an important part of social interaction, and frequent misunderstandings negatively impact self-confidence, social participation and overall well-being across the lifespan \citep{hoppe_sprachverstehen_2014, punch_social_2011, hogan_higher_2015}. For hearing-impaired listeners, conversations in groups are typically challenging, especially when background noise is present. Hearing devices aim to increase speech intelligibility and reduce listening effort, but about 25\,\% of users continue to report low satisfaction with their device in difficult listening situations, according to large-scale surveys \citep{carr_marketraktracking_2022, kim_factors_2022}. Especially for patients with higher degrees of hearing loss, current hearing devices provide insufficient support \citep{hoppe_sprachverstehen_2014, punch_social_2011}.\\
Improving device performance is not simple, as it requires identifying and attenuating irrelevant signals in diverse and dynamic sound scenarios. This task may be facilitated by recent developments that use the user's gaze and head movement behaviour to distinguish between relevant and irrelevant sources \citep{Best2017,FavreFelix2018,Grimm2018}.
To evaluate such behaviour-based signal processing strategies, adequate methods are required. 
For example, interactive conversation, in contrast to isolated listening, was found to elicit different head movement behaviours from the listener \citep{hadley_speech_2019, hartwig_speaking_2021}. 
Therefore, systematic tests that elicit natural behaviour are needed to evaluate hearing device performance. This is suggested in particular by evidence that widely used evaluation methods and measures, such as speech reception thresholds, are poor predictors of device performance in real life when used to evaluate more complex algorithms \citep{naylor_theoretical_2016, keidser_introduction_2016}. This may be partly due to the fact that the head movement behaviour in these traditional methods does not reflect real life behaviour.\\
Virtual audio-visual environments are a versatile tool for presenting everyday life scenarios. In contrast to field measurements, the presented stimuli can be precisely controlled, multiple scenarios can be presented in one location, and target and noise signals can be directly accessed. Recently, the importance of non-verbal behaviour in virtual conversations has been emphasised, in particular the importance of head movement and eye contact for effective communication and comfort \citep{rogers_realistic_2022, aburumman_nonverbal_2022}. Regarding self-motion, \citet{hendrikse_influence_2018} showed that the listener's head movement behaviour depends on the level of lip movement and head orientation animation of virtual animated characters when following a conversation. It was also found that manipulating the head movements of an interlocutor in an active conversation also changes the movements of the receiver \citep{boker_effects_2009}. An earlier overview \citep{Whalen2003} concludes that using captured behavioural data will achieve a higher realism of avatars. 
Therefore, when assessing the listener's behaviour, it is crucial to accurately represent potential interlocutors in a virtual environment in terms of non-verbal cues such as head orientation, gaze behaviour and nodding.

This work aims to contribute to current research by investigating the effects of head movements of avatars on the listener's behaviour and experience. The research question of this study is:
\begin{center}
\textit{Does the transmission of head movements to
avatars affect behaviour and experienced involvement in
virtual environment scenes?}
\end{center}
Our approach was to observe participants in real interactive triadic conversations in telepresence, and systematically modify the head movement animation level of avatars representing the interlocutors. Avatars were displayed using virtual animated characters. Difficulty was controlled by changing the background noise level. The effect of head movement animation level was evaluated using both objective behavioural data and subjective experience ratings. The measurement environment represented a typical conversation scenario in a pub.

We expected that the head movements of the interlocutors would have an influence on the communication behaviour and conversation success, as they contain non-verbal communication cues \citep{aburumman_nonverbal_2022}. 
One aspect that can reflect a change of effort during conversation is speech behaviour. 
A typical compensation strategy during spoken interaction in noise is a shift in speech level and spectral content, known as the Lombard effect \citep{brumm_evolution_2011}.  
As a simple measure of the Lombard effect, an analysis of \textbf{speech level} is possible, as was done for face-to-face conversation by \citet{hadley_speech_2019, hadley_conversation_2021}.  
In addition, it was found that \textbf{utterance duration} was shorter in high background noise during free conversation, presumably to simplify information \citep{hadley_speech_2019, hadley_conversation_2021}. In contrast, utterance duration was found to increase in a dyadic puzzle task \citep{beechey_measuring_2018}. 
Additionally, the timing of switches between successive talkers also may indicate the ease of interaction during a conversation. The impaired alignment of speech contributions is sometimes associated with reflecting required speech processing time. However, effects are multi-factory and ambiguous \citep{heldner_pauses_2010, hadley_speech_2019}. For example \cite{hadley_speech_2019} found shorter speech gaps between successive talkers in higher noise levels, although the presence of noise was expected to lead to increased speech gaps due to higher speech recognition effort.
Previous studies use various terminology, such as `inter-speaker pauses' and `inter-speaker overlaps' \citep{hadley_speech_2019} or the `floor transfer offset' \citep{heldner_pauses_2010, levinson_timing_2015}. In this work, pauses between two talkers are referred to as \textbf{speech gaps} and overlapping speech at speaker turn takes as \textbf{speech overlap} (cf. Figure~\ref{tab:ttscenarios}), following the terminology of \citet{heldner_pauses_2010}.  

Next, changes in background noise as well as avatar head movements were expected to affect participants' movement behaviour (translation and head orientation). Changes in head movement behaviour are relevant regarding possible implications for head-worn hearing devices.  For in-person triadic conversations, \citet{hadley_conversation_2021} did not find an affect of noise level on head orientation angle, but a larger head orientation angle in babble compared to speech-shaped noise.
\citet{lu_investigating_2021} found a small effect of noise level for young normal-hearing listeners, but not for older hearing-impaired listeners. Yet, since this study also varied visual aspects of the conversation, the \textbf{angular distance} to interlocutors might be affected by animation conditions. Additionally, the \textbf{head orientation range} during one condition can be calculated as a measure of head movement behaviour. This represents the range between the two most prominent orientation angles, which were expected due to the two other interlocutors. 

For the collection of subjective \textbf{experience ratings} of participants, several possible questionnaires are available.
The igroup presence questionnaire (IPQ) \citep{schubert_experience_2001} evaluates the experienced sense of presence, with the factors spatial presence, involvement and realism of a virtual environment.
This questionnaire was also used in a previous study \citep{hendrikse_movement_2019} to evaluate non-interactive virtual scenes of everyday life, using the same virtual animated characters as used in the present study.
When evaluating an interactive communication scenario, the perceived success of the conversation is an important aspect. \citet{nicoras_conversation_2022} identified seven clusters of statements describing a successful conversation, from normal-hearing and impaired-hearing listeners. They include passive and active participation aspects in one-to-one or group conversations.

Our general expectations were to find an effect of background noise on the selected measures compared to a quiet condition, if the noise level was set to a sound pressure level that considerably impeded conversation. Different implementations of head movement were expected to have smaller effects, as a normal-hearing participant group may rely less on visual compared to auditory cues. Possibly, effects may be present only in noise, where the acoustic interaction is limited and visual communication may be more relevant.

Following the presentation of the experimental paradigm, apparatus and methods utilised in this study, the results are presented and discussed in the context of recent literature and possible implications for the development of evaluation methods, hearing device algorithms and clinical research.


\section*{Methods}

\subsection*{Design and task}
Free interactive triadic conversations were conducted in virtual reality using telepresence technology. The conversation scenario consisted of three interlocutors, the test participant and two confederates, seated at equal distances around a virtual table (cf. Figure \ref{fig:triangle}).  Triadic conversations are the minimal configuration in which turn-taking potentially results in head movement and they are a common scenario in everyday life.  For the participants, the interlocutors (the two confederates) were visually represented in the lab by avatars consisting of virtual animated characters with different levels of head movement animation, or by videos. For the confederates, the participant was visually represented over video. The virtual environment was set in a virtual pub and projected onto a screen. This setup facilitated the manipulation of the interlocutors' head movements. All three interlocutors were seated in different rooms, which allowed access to separate speech and noise audio signals.

Only one participant was invited to each session. The other two interlocutors, the confederates, took part in several sessions. They also met the requirements of the study participants. One avatar always represented one person (labelled \textit{Mar}) and the other avatar (labelled \textit{Joa}) represented two different people, each taking part in about half of the measurements. One confederate (\textit{Mar}) controlled the ongoing experiment and was therefore not blind to the measurement conditions. This person is also referred to as the experimenter.

To achieve active participation, participants were instructed to maintain a conversation about casual topics. Picture cards, as suggested by \cite{Smeds2021} and also used by \cite{hartwig_speaking_2021}, were used to spark the conversation topics, which could freely change, or span multiple conditions. The picture cards were given only to the confederates to keep the focus of participants on the avatars.

For a detailed evaluation of avatar head movement effects, multiple levels of head movement realism were implemented together with two contrasting levels of background noise, resulting in a 2$\times$4 factorial design, see Table~\ref{tab:conditions}. In the visually static condition ('stat'), the avatar's head and gaze were fixed over the centre of the table. In the automatic head movement condition ('auto'), the avatar reactively oriented the head and eye gaze towards the defined target angle, cued by speech onset detection \citep[Figure~2]{hendrikse_influence_2018}. To include a more complex behaviour in one condition, the confederate's head movements were recorded by a motion sensor, transmitted to the visual simulation, and used to control the avatar's head ('trans'). In the condition which was considered to be closest to a face-to-face conversation, a video image of the confederate's head and shoulders was transmitted to the position of the avatar in the virtual scene ('video').

\begin{table}[h]
  \includegraphics[width=\columnwidth]{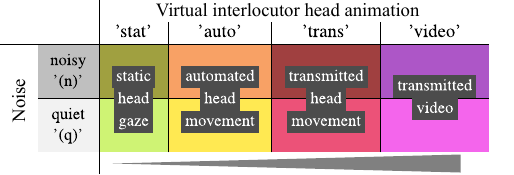}
  \caption[Selected independent measurement variables and levels]{\footnotesize Selected independent variables and levels of avatar's head movement (\enquote{animation}) and background noise (\enquote{noise}) in a 2×4 factorial design. The cone indicates the qualitative magnitude of the expected level of realism.}
        \label{tab:conditions}
\end{table}

Each audiovisual condition lasted at least 5 minutes and was manually switched to the next condition by the experimenter at an appropriate time to avoid abrupt interruptions of the conversation. The order of conditions was pseudo-randomised in advance. After each condition, participants completed a questionnaire in which they rated their experience during the last conversation.

\subsection*{Measurement setup}

The measurement setup was distributed over three rooms: one for the participants with a audiovisual reproduction using loudspeakers and a cylindrical video projection, and one for each confederate, with binaural headphone reproduction and video reproduction on computer screens.

\subsubsection*{Reproduction setup}
Participants were seated in the main laboratory, surrounded by 45 full-range loudspeakers (Genelec 8020D; 16 at ear level on a circle with a diameter of 3.7~m, 6 at $-30\degree{}$ elevation, 16 at $+15\degree{}$ elevation, 6 at $45\degree{}$ elevation and one at $90\degree{}$ elevation) and a $300\degree{}$ field of view video projection (three NEC U321H, at $-105\degree{}$, $0\degree{}$ and $105\degree{}$, each with a field of view of $120\degree{}$ and a blending overlap of $15\degree{}$) on a cylindrical, acoustically transparent projection screen \citep{Heeren2023}. Acoustic and visual rendering were performed on separate computers connected via a network. The audio system was run at a sampling frequency of 48~kHz using block-based processing with a block size of 128 samples for seven participants and 256 samples for nine participants (the block length was increased to avoid artefacts due to dropouts). The Toolbox for Acoustic Scene Creation and Rendering (TASCAR, version 0.228, \cite{grimm_toolbox_2019}) was used to reproduce the virtual acoustics, with vector-based amplitude panning \citep{Pulkki1997} used for rendering to loudspeakers. The visual model was reproduced with the Blender game engine (version 2.79b, \cite{roosendaal_blender_2017}), with a virtual camera at the position of the participant. The blending and warping was processed on the graphics card (NVIDIA Quadro M4000). To increase immersion in the virtual environment, the participant's head position was tracked and the participant was allowed to move in the audiovisual environment by updating the horizontal position of the receiver and camera in the audiovisual simulation. To minimise the number of sensors on the participants, their speech was recorded using a directional room microphone (Sennheiser MKE600) installed at the top of the projection screen. In order to maximise the signal-to-noise ratio of the microphone signal and minimise echoes from the remote interlocutors, the transfer functions from each loudspeaker to the room microphone were recorded for static echo cancellation before each session.

The two confederates were seated in separate rooms. Each confederate wore a headset (AKG HSC 271) equipped with a head tracking sensor, see section \ref{sec:sensors}{\em Motion capture} for details. The speech signal of the confederates was picked up by a headset microphone and high-pass filtered ($f_{cut}\!=\!120$~Hz) to compensate for the near-field effect. The confederates saw the participant and each other via cameras on a computer screen (Dell P2422H, 0.61~m screen diagonal), one for each person. Web cameras (Logitech C270 HD) were mounted on the computer screen or centred between the avatars so as not to cover them. The arrangement of computer screens formed an equilateral triangle between the interlocutor and the two computer screens to represent the viewing angles and to elicit correct head orientation angles in the interlocutors. The experiment was controlled via a graphical interface on a laptop computer by one of the confederates, the experimenter.

%
The main laboratory was acoustically treated with absorbers on the ceiling and walls, a carpet on the floor and a heavy stage backdrop around the spherical loudspeaker array. The resulting reverberation time was below 0.2~s for all frequencies above 500~Hz and below 0.4~s for 125 and 250~Hz. The Direct-to-Reverberant Ratio was increasing from -3.7~dB at 125~Hz to 8.9~dB at 4~kHz, measured from the frontal reproduction loudspeaker at the listening position. One of the confederates' rooms was another acoustic laboratory and the third was an office.

\subsubsection*{Acoustic calibration}
The loudspeakers were calibrated with pink noise in the frequency range 80~Hz to 10~kHz. The frequency response of the loudspeakers and the room, measured in one-third octave bands and averaged over three bands, was calibrated by fitting four second-order parametric equalisers. The subjects' headphones were calibrated by subjective comparison of a headphone-reproduced sound source with a real physical source. The real and simulated sources were placed at a distance of 1.2~m from the listener and played a pink noise in the frequency range 80~Hz to 10~kHz. The headset microphone was calibrated by achieving the same sound pressure level at the participant's position in the laboratory at a distance of 1.2~m using continuous spontaneous speech.

\subsubsection*{Motion capture}\label{sec:sensors}
The participant’s head translation and rotation in three dimensions was tracked with a reflective marker crown, captured by six infrared cameras (Qualisys Miqus M3, Qualisys Track Manager). The confederates' head rotation in three dimensions was recorded by an inertial measuring unit consisting of a gyroscope, an accelerometer and an Arduino processor, which was mounted on the headset. All movement data were tracked with a sampling rate of $f_s\!=\!100$~Hz.

\subsubsection*{Network transmission and data recording}
The distributed setup was realised by transmitting audio, video and motion data over the local university network. The `OVBOX' system \citep{grimm_orlandoviols_2021,Grimm2024} was used for audio and motion transmission and experimental control. It provided the transfer of audio data between interlocutors, and transmitted head movement data as well as control commands to the acoustic rendering toolbox. Audio data were then included into the scene, head movement data were forwarded to the visual scene, and commands controlled the virtual scene. As an example, for each confederate, the audio signals of the room microphone and the other confederate plus noise from the virtual scene were transmitted uncompressed via network, rendered locally considering the spatial distribution and orientation of interlocutors, and played back over headphones. This setup also enabled a remote control of the experiment by one confederate.
The `OVBOX' system includes a central relay server, which was also running within the university network. In the `OVBOX' system, data is typically transferred using the Universal Datagram Protocol (UDP). Here, a Transmission Control Protocol (TCP) connection was used between the first confederate (experimenter) and the relay server, at the cost of a greater delay, to ensure secure transmission of all control information without potential packet loss. A peer-to-peer live video transmission service \citep{steve_seguin_vdoninja_2022} was used for video transmission.

The audio delay from the interlocutor microphone to the position of the participants was 49.8\,ms, at a block size of 256 samples via the Transmission Control Protocol (TCP). This delay is larger than typical delays reported by \citet{Grimm2024}, which is caused by the relatively large block size and the TCP transmission.
The delay between the remote head movement and a movement of the projected image of the avatar was approximately 180\,ms. The streamed video pictures had a delay from camera to recipient browser of roughly 221.7\,ms and an additional delay from the browser picture via Blender onto the lab screen of 290.2\,ms.

Data from the rating questionnaire was collected via a tablet and sent to the data logger. Short term RMS speech levels were measured in each audio block and sent to the data logger. Data collection was centralised on the audio rendering PC in the main laboratory using the data logging module in TASCAR. This allowed all data sources to be synchronised.

\subsection*{Virtual environment}
The virtual space represented a pub with other guests at several tables in the background, see central panel of Figure \ref{fig:triangle}. The dimensions, acoustic characteristics and audiovisual model are based on an existing room, the ``Ols Brauwerkstatt'' in Oldenburg, Germany \citep{grimm_giso_pub_2021}.
The participants were located at a round table together with two other interlocutors at approximately equal distances ($d_{mean}\!=\!1.12\,\mathrm{m}\pm 0.03\,\mathrm{m}$) to form an equilateral triangle.

\bgroup
\begin{figure}[!htbp]
\centering
\includegraphics[width=0.7\linewidth]{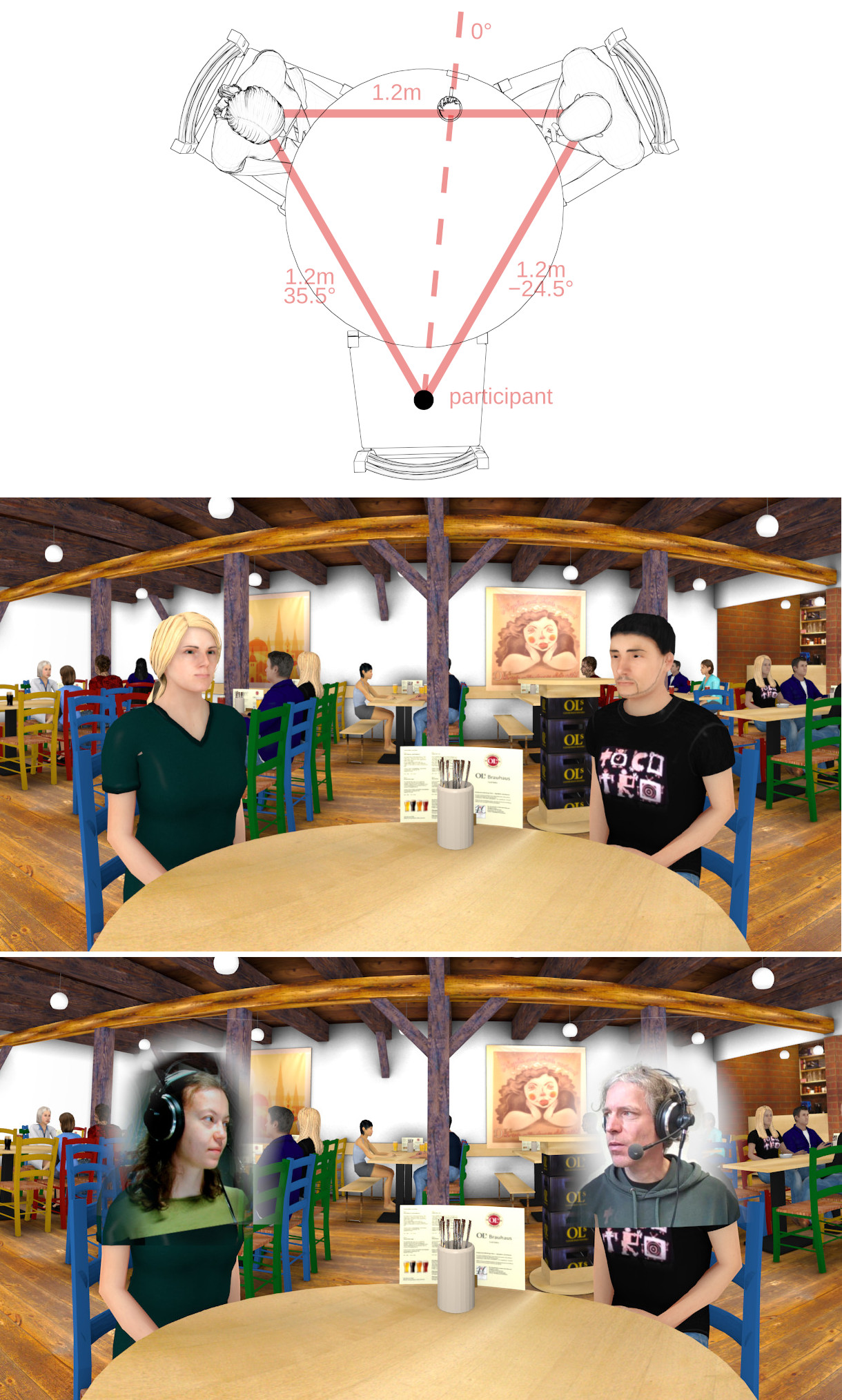}

\caption{\footnotesize Top panel: top view of the virtual conversation setup. The test participant is virtually sitting on the empty chair. Centre panel: view of the virtual environment from the participant's perspective. Bottom panel:  view of the virtual environment with real-time video textures in front of the avatars' faces, which was used in the condition 'video'.}
\label{fig:triangle}
\end{figure}
\egroup

In the acoustic model, the interlocutors were virtually represented by omnidirectional point sources. Noise sounds were first-order ambisonics field recordings of a babble noise in the local university cafeteria and a refrigerator sound \citep{grimm_first_2019}, and  music \citep{dokapi_lighthouse_2015}.  
The noise sounds were implemented as diffuse sound fields, and the music played through two loudspeakers placed in the room with a physical model based loudspeaker cabinet simulation of a loudspeaker with a resonance frequency of 80~Hz, see the documentation of the module ``spksim'' in \citet{GisoGrimm2022} for details.
In the 'quiet' conditions, the refrigerator sound ensured a steady noise floor level at 37.3~dB~SPL~(A) (48.2~dB~SPL~(C)) to mask the varying ventilation noise of the projectors. In 'noise', the  sounds fridge, babble and music resulted in an $L_{eq}$ of 69.1~dB~SPL~(A) (71.1~dB~SPL~(C)) at the listening position. 

In the visual simulation of the virtual environment, the avatars were blinking at random times and moved slightly back and forth to indicate breathing. Their eye gaze was either directed horizontally over the centre of the table (condition 'static') or towards the other avatar's or participant's head (conditions 'auto' and 'transmitted'). The gaze flipped when the avatar's head crossed the centre angle between the two interlocutors. This avoided confusion, but transferred no additional information. When the confederates were speaking, the avatar's lips were moved via a speech-based algorithm in real-time \citep{llorach_web-based_2016}. The virtual characters had no facial expression, gestures or body movement.  For the 'video' condition, a real-time video of the confederate's shoulders and head was streamed to the blender scene and shown on a flat texture in front of the avatar's face, see lower panel of Figure \ref{fig:triangle}.  The virtual characters in the background were only visible during noisy conditions.

\subsection*{Participants}
Sixteen participants (nine female, seven male) were acquired via a university-wide online advertisement. The study was approved by the Commission for Research Impact Assessment and Ethics of the Carl von Ossietzky University of Oldenburg (approval number EK/2021/068), and participants were reimbursed for their time. Their age ranged from 19 to 32 years (mean 23.8 years). They stated to have no hearing impairment and  language skills at or close to a native speaker level in German, the language of conversation. Vision was corrected to normal if necessary. 14 out of 16 persons were not acquainted to at least one of the confederates. 

\subsection*{Measures and analyses}\label{sec:Analyses}
One data point per participant and condition was calculated for each selected measure. Variance within conditions was not analysed, except for the head orientation measure, where range was used as a separate measure. Table~\ref{tab:measures} summarises the selected measures evaluating the dependent variables.

\begin{table}[h]
  \raggedright
\footnotesize
\begin{tabular}{|l|l|c|}
  \hline
  \multirow{2}{*}{\bf Variable}  & \multirow{2}{*}{\bf Measure}   & \multirow{2}{1cm}{\bf Sensor / Question} \\
  \\
  \hline
  \hline
  \multicolumn{3}{|l|}{Communication behaviour:}                                        \\
  \hline
      \multirow{4}{1.5cm}{speech behaviour}    & speech level                          & mic          \\
      { }               						& utterance duration                    & mic          \\
                                 				& speech gap duration                   & mic          \\
                                 				& speech overlap duration               & mic          \\
  \hline
      \multirow{3}{1.5cm}{head movement behaviour}	  & angular distance to speaker     & mocap        \\
      { }                						& head orientation range                & mocap        \\
      { }               						& head translation                      & mocap        \\
      \hline
      \hline
  \multicolumn{3}{|l|}{Experience:}                                                     \\
  \hline
      \multirow{6}{1.5cm}{sense of presence}& presence in virtual space  & Q1           \\
      {}                 		  & copresence with interlocutors         & Q2           \\
                                 & lack of awareness of real environment & Q3           \\
                                 & realism of virtual world              & Q4           \\
                                 & realism of avatars                    & Q5           \\
                                 & realism of voice                      & Q7           \\
  \hline
  \multirow{4}{1.5cm}{conversation success}  & being spoken to in a helpful way      & Q6           \\
  { }                   		  & being able to listen easily           & Q8           \\
                                 & sharing information as desired        & Q9           \\
                                 & making an effort to be understood     & Q10          \\
  \hline
\end{tabular}
\caption{\footnotesize Overview of the dependent variables and corresponding measures. Behaviour was assessed using physical sensors: speech behaviour was based on level analysis of head-mounted or room microphones (mic), and movement was measured using an optical motion capture system (mocap). A questionnaire was used to quantify the sense of presence and the conversation success, see Table~\ref{tab:questionnaire} for the presented questions (Q1...10) and their translations.}
\label{tab:measures}
\end{table}

\subsubsection*{Speech behaviour}\label{sec:AnalysesSpeech}
\begin{figure}[h]
  \centering
  \includegraphics[width=\columnwidth]{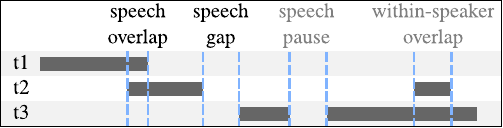}
\caption{\footnotesize Scheme of possible combinations of speech segments in a triadic conversation. Each row represents one speaker, the dark grey bars are speech activity over time. Gaps and overlaps are marked by light blue bars. Only the times of speech gaps and overlaps were defined as turn takes.}
\label{tab:ttscenarios}
\end{figure}

The speech analysis was based on short-term levels, which were calculated as mean-square in blocks of 2.9~ms or 5.8~ms length (after 7 measurements, the block length was increased to avoid artefacts due to dropouts). These short-term levels were smoothed with a 10~Hz first-order Butterworth low-pass filter and then converted to dB SPL. The two block lengths correspond to a sampling rate of approximately 344~Hz and 172~Hz, respectively. Both sampling frequencies are expected to provide sufficient temporal resolution.
As in \citet{hadley_conversation_2021}, short-time levels were smoothed with a rolling 100\,ms Hann window. For a robust speech activity detection, a noise floor estimation  
was done by fitting a linear model to the level distribution. The level thresholds for speech activity were set to the 25\,\% point of the estimated dynamic range above the noise floor for every person and each condition individually. Speech activity was then defined as all speech levels above threshold. If there was a pause between sections shorter than 1.25\,s, the two sections were connected, following \citet{hadley_conversation_2021}. All activity sections shorter than 0.05\,s were discarded. \\ 
The average \textbf{speech level} was calculated by taking the 95th percentile of speech levels during detected speech activity for each participant. \textbf{Utterance duration} was the median duration of all speech activity segments. 
\textbf{Speech gap duration} and \textbf{speech overlap duration} were based on the interaction of interlocutors, i.e. based on the time-aligned speech level data, and defined according to \cite{heldner_pauses_2010}, see Figure~\ref{tab:ttscenarios}. At speaker turn takes, speech activity pauses or simultaneous activity were detected and the median duration of the respective segments was calculated.
The speech overlap duration was noted in negative values, speech gap duration at turn takes in positive values. 

For the analysis of head movement behaviour during the conversation, segments of listening and segments of speaking were distinguished for each interlocutor. Speaking was defined as the period of speech activity during which only the respective interlocutor was active, in order to avoid ambiguous data. It was assumed that the interlocutors were listening during the whole condition, except when they were speaking themselves. This implies that participants were also attentive during speech pauses.

\bgroup
\begin{figure*}[phtb] 
\centering \makeatletter
\includegraphics[width=0.8\linewidth]{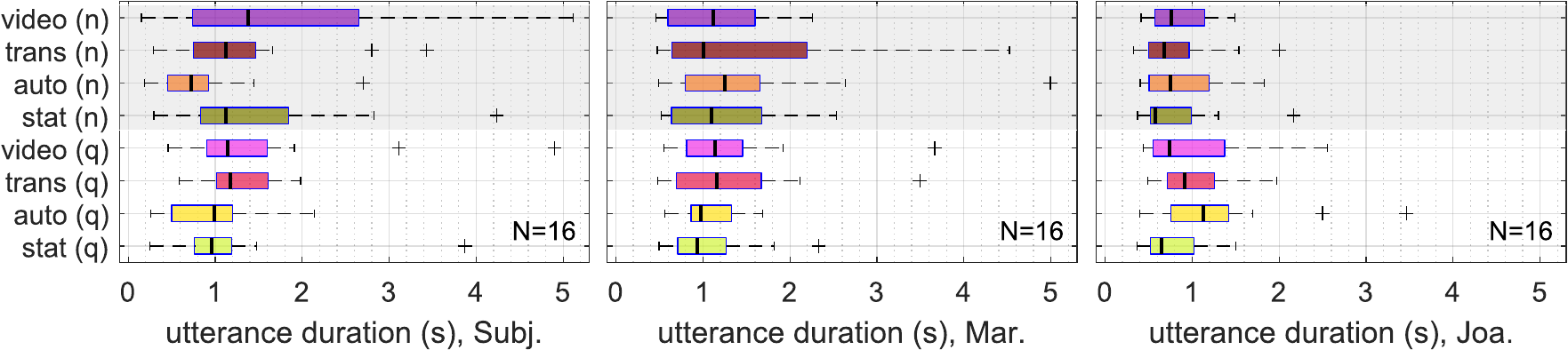}
\makeatother 
\caption{\footnotesize Median utterance duration during conversation of participants \textit{Subj} and confederates \textit{Mar} and \textit{Joa}. Note that only participants, not the two confederates, were exposed to different visual conditions. 
Four outliers are outside
the displayed x value range. \textit{Mar}: in trans (n) $6.46\mathrm{\,s}$, in stat (n) $7.83\mathrm{\,s}$, \textit{Subj}: in trans (n) $6.36\mathrm{\,s}$, in stat (n) $7.33\mathrm{\,s}$.
}
\label{fig:speech_utt_dur}
\end{figure*}
\egroup

\subsubsection*{Head movement behaviour}
The \textbf{angular distance} to the speaker was defined as the median value of absolute angle distances between the participant's head rotation angle and the angle towards the centre of the avatar's face around the z-axis (i.e. the head yaw angle).  
Only times during the respective avatar was speaking were considered. The reference points for the avatars' faces were located between their eyes and mouth, derived from their position in the condition 'static'. For an overall measure representing both avatars, the median yaw angle to the corresponding avatar over all conditions was subtracted from the angular distances before calculating the cumulative median. This was done to account for a systematically smaller angular distance to the avatar \textit{Mar} by most participants.
The \textbf{head orientation range} was represented by the difference of the 5th and 95th percentile of yaw angles. This was calculated separately for times during listening, and speaking (definition see previous section).  
To find individual behavioural changes, the difference of absolute values to the median over all conditions was calculated.
The \textbf{head translation} of participants was calculated as the median position in the direction towards the centre between the two virtual characters. Here again, the difference of absolute values to the median over all conditions was calculated to find individual changes.

\subsubsection*{Experience rating} 
The \textbf{rating questionnaire} was a set of 10 questions on a 7-point rating scale with verbal anchors at the floor, centre, and ceiling value. Six questions were inspired by the igroup presence questionnaire (IPQ, \cite{schubert_experience_2001}) which evaluates the experienced presence, involvement and realism of a virtual environment. Four items were questions based on the clusters of conversation success identified by \citet{nicoras_conversation_2022}. The questions were phrased in English based on the proposed clusters and then translated into German. The items of the questionnaire are noted in Table~\ref{tab:questionnaire}.
The questionnaire therefore collected one rating value from the participants for each question and condition. The resulting data points were not further pooled or processed before the statistical analysis.

\subsubsection*{Statistical analysis}
A two (noise level) by four (animation level) repeated-measures univariate analysis-of-variance (ANOVA) was performed for every dataset. Sphericity was tested with Mauchly’s Test of Sphericity, and in the case of violation the degrees of freedom were corrected with a Greenhouse-Geisser correction factor. 
A main effect was significant if $p\leq\alpha$, with $\alpha = 0.05$. A possible main effect of \enquote{animation level} or an interaction effect was further evaluated with a pairwise comparison of the estimated marginal means to obtain the differences between individual levels. The inflation of the family wise error rate was controlled via the Bonferroni correction as a conservative control over the false rejection of a true null hypothesis. Additionally, the effect size of significant effects was estimated by calculating the partial eta $\eta^2_p$. If a measure was analysed for all three interlocutors, \textit{interlocutor} was included as a between-subjects factor to find differences between the interlocutors.

    
\section*{Results}
The study data consists of 16 sessions, with one missing session for head movement data due to a technical error. The participant group data are indicated by \textit{Subj}, whereas the two confederates in the triad are indicated by \textit{Mar} and \textit{Joa}. Keep in mind that the confederates were not exposed to the visual conditions. \textit{Mar} consists of multiple measurements of one person and \textit{Joa} of two  alternating persons. 

\subsection*{Speech behaviour}

The effect of \enquote{noise} on participants' speech level was significant, see Table~\ref{tab:effectmeasures}, 
with an mean increase in noise of $10.6$~dB. 
Participants also showed a slightly higher speech level with a higher level of head movement animation, which was significant as a main effect, but not significant for any pairwise comparison due to the Bonferroni correction of $p$. When \textit{interlocutor} was used as a between-subjects factor, the three interlocutor datasets were not found to be different, except that \textit{Joa} had a slightly lower speech level than \textit{Subj} ($p_{\mathrm{bonf}}\!=\!0.047$, mean difference: $1.8$~dB). 
In low background noise, the median speech level was at $67.6$~dB~SPL (\textit{Subj}), $65.1$~dB~SPL (\textit{Mar}) and $65.1$~dB~SPL (\textit{Joa}).  
In noisy conditions, median values increased to $78.2$~dB~SPL (\textit{Subj}), $79.2$~dB~SPL  (\textit{Mar}) and $77.2$~dB~SPL (\textit{Joa}).

The median duration of utterances during conversation are shown in Figure~\ref{fig:speech_utt_dur}. Pooled across all conditions, the median duration was $1.08$~s (\textit{Subj}), $1.15$~s (\textit{Mar}) and  $0.71$~s (\textit{Joa}).
For the participants, utterance duration was affected by \enquote{animation}, see Table~\ref{tab:effectmeasures}, 
with longer utterances in 'video' compared to 'automatic' (
mean difference: $0.79$~s). There was no significant effect of \enquote{noise} %
and no interaction effect ($p\!=\!0.404$). \textit{Joa} showed slightly shorter utterance duration in noise (mean difference: $0.21$\,s), $F(1,15)\!=\!9.27,\, p\!=\!0.008,\, \eta^2_p\!=\!0.382$. When \textit{interlocutor} was used as a between-subjects factor, \textit{Joa} had shorter utterances than \textit{Mar} ($p_{\mathrm{bonf}}\!=\!0.047$, mean difference: $0.396\mathrm{\,s}$) and \textit{Subj} ($p_{\mathrm{bonf}}\!=\!0.038$, mean difference: $0.428\mathrm{\,s}$). \textit{Mar} and \textit{Subj} were not found to be different.

\bgroup
\begin{figure}[phtb]
\centering \makeatletter
\includegraphics[width=1\linewidth]{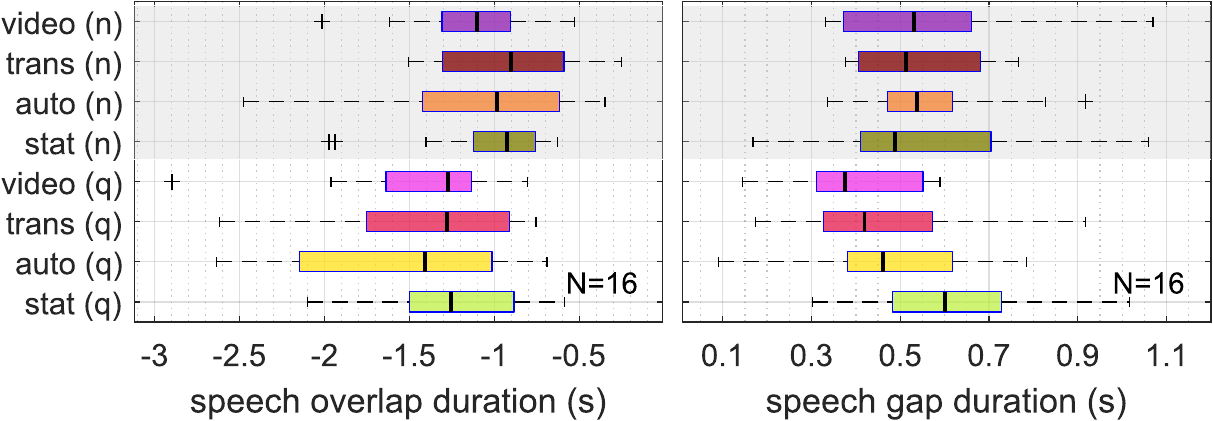}
\makeatother 
\caption{\footnotesize Median duration of speech overlaps (left) and gaps (right) at turn takes between the three interlocutors. The speech overlap duration is noted in negative values. Note that the frequency of overlaps decreased in noise, whereas the frequency of gaps increased (Not shown here). Right panel: One outlier is outside the displayed x value range (in auto (n): 1.54\,s). 
}
\label{fig:speech_gapoverl}
\end{figure}
\egroup

The median duration of speech gaps and overlaps at turn takes are shown in Figure~\ref{fig:speech_gapoverl}. The median duration of all detected speech gaps was $0.51$~s, speech overlaps had a median duration of $1.11$~s. 
ANOVA results showed that speech gaps and overlaps were significantly affected by \enquote{noise}, but not by \enquote{animation}, see Table~\ref{tab:effectmeasures}. Speech gap duration increased in noise by $0.074$~s in mean value. 
A possible interaction effect (see Figure~\ref{fig:speech_gapoverl}) was not significant ($p=0.066$).
Additionally, it was found that in noisy conditions, speech gaps occurred more often, with a median increase of $\Delta f_{\mathrm{gaps}}=0.012$~Hz, which corresponds to $+3.6$~gaps during a condition duration of $300$~s.
Speech overlaps were shorter in noise (mean difference: $0.36$~s). 
The number of overlaps was reduced in noise, with a median frequency difference of $\Delta f_{\mathrm{diff,overl}}\!=\!-0.012\mathrm{\,Hz}\!=\!-3.6\mathrm{\,gaps/}300\mathrm{\,s}$. 

\subsection*{Head movement behaviour}

\bgroup \begin{figure*}[phtb]
  \centering \makeatletter
  \includegraphics[width=\textwidth]{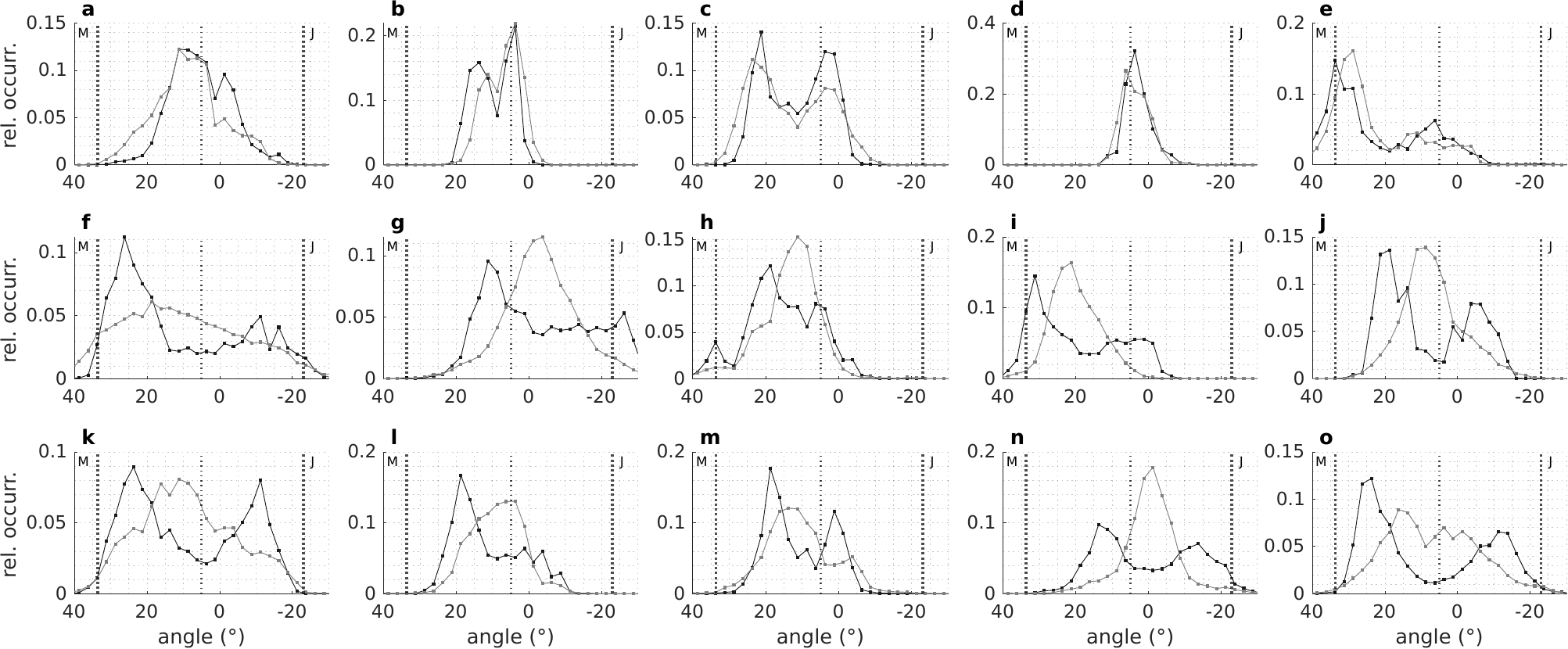}
  \makeatother 
  \caption{\footnotesize Relative occurrence over head yaw angle for each participant during listening (dark grey) and speaking (light grey), pooled over all conditions. The vertical lines indicate the position of the avatar \textit{Mar} (M), the avatar \textit{Joa} (J), and the centre between both. The avatars' positions are symmetric around 5 ° in lab coordinates. Occurrence values sum up to 1, and the angle is divided into increments of 2.5 °. Definition of listening and speaking: see section~\nameref{sec:Analyses}.}
  \label{fig:movem_indiv}
\end{figure*} \egroup

Participants consistently did not orient their heads completely towards the avatars' faces, but showed some amount of head angle undershoot. In Figure~\ref{fig:movem_indiv}, individual histograms of head movement angles of every recorded session are displayed. Visually different patterns were found between listening and speaking, mostly pronounced in individuals (f) to (o).

The overall median angular distance was not symmetric, participants were more closely oriented towards the avatar \textit{Mar} ($15.4\,^\circ$, median over all conditions and participants) compared to avatar \textit{Joa} ($23.4\,^\circ$). The pooled data over both avatars, see Figure~\ref{fig:movem_angle_dist}, showed that angular distance changed with levels of head movement animation and in noise. 
Participants oriented their heads more accurately towards the avatar faces in noise by a mean of $1.6\,^\circ$ and had a closer orientation in 'video' compared to 'static' by $2.1\,^\circ$, see Table~\ref{tab:effectmeasures}.

\bgroup
\begin{figure}[phtb]
\centering \makeatletter
\includegraphics[width=0.575\linewidth]{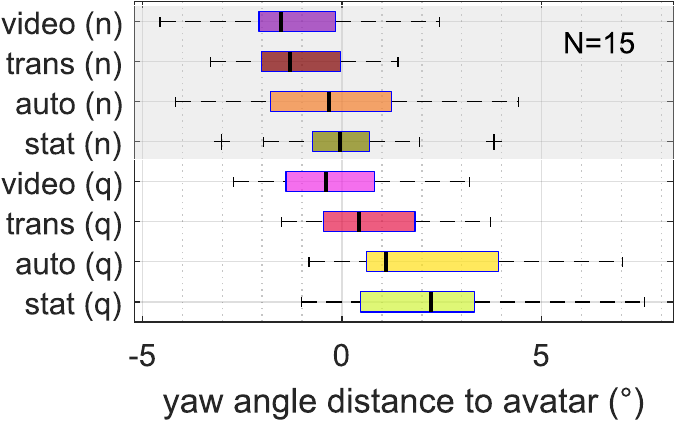}
\makeatother 
\caption{\footnotesize Participant's angular distance to the avatars’ faces in each condition, referenced to the individual median over all conditions. Lower values represent a more accurate orientation. One outlier is outside the displayed x value range (in video (q): -11.48\,°).}
\label{fig:movem_angle_dist}
\end{figure}
\egroup

The used range of head yaw angles was individually different (cf. Figure~\ref{fig:movem_indiv}), individual median values over all conditions during listening or speaking ranged from $11.2\,^\circ$ to $48.3\,^\circ$, with a median of $28.2\,^\circ$ over all participants. The separate analysis during speaking and listening resulted in different effect patterns, see Figure~\ref{fig:movem_angle_range}.
The ANOVA resulted in a significant main effect of \enquote{animation} on the range of head yaw angles during speaking, but not during listening. In noise, head orientation range was increased during speaking and listening, see Table~\ref{tab:effectmeasures}. 

\bgroup
\begin{figure}[phtb]
\centering \makeatletter
\includegraphics[width=1\linewidth]{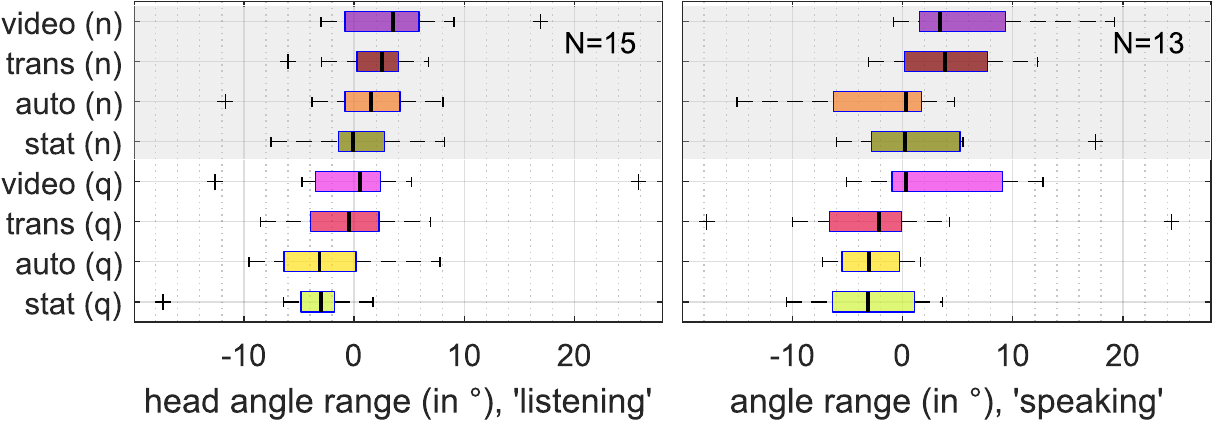}
\makeatother 
\caption{\footnotesize Head orientation range during ‘listening’ (left) and ‘speaking’ (right), referenced to the individual median over all conditions. 
 A higher value represents a wider angle range. Two data sets had a 'speaking' period below 15\,s in at least one condition and had to be excluded.}
\label{fig:movem_angle_range}
\end{figure}
\egroup

The head translation towards the screen which displayed the two interlocutors was affected by noise, see Table~\ref{tab:effectmeasures}. 
Participants positioned their heads $3.0\mathrm{\,cm}$ closer to the screen in noisy conditions. There was no significant main effect of \enquote{animation} 
and no interaction effect ($p\!=\!0.366$).

\subsection*{Experience rating}
Participants rated their presence in the virtual space (Q1, see Table~\ref{tab:questionnaire}) and their co-presence with the interlocutors (Q2) with a median value of $+2.0$\,(interquartile range $\mathrm{IQR}\!=\!1.0$), where $+3.0$ corresponded to 'completely true' and $0$ to 'neither nor'. The rating was not affected by \enquote{animation} or \enquote{noise}, see Table~\ref{tab:effectmeasures}. 
They indicated to be less aware of the lab environment (Q3) in the 'video' condition compared to 'static'. 
The realism of the scene (Q4) was rated with a median value of $+1.0\,(\mathrm{IQR}\!=\!2.0)$, with slightly higher ratings in noise compared to quiet. 
The avatars were rated as more visually realistic (Q5) with higher levels of \enquote{animation},
resulting in a significant difference of $+1.9\mathrm{\, points}$ between 'static' and 'video' 
and a close-to-significant difference between 'static' and 'transmitted'. 
The voice of the interlocutors (Q7) was rated consistently as realistic (median: $+3.0,\,\mathrm{IQR}\!=\!1.0$).

For all four questions regarding conversation success, the effect of noise was significant. Participants indicated that they were less spoken to in a helpful way (Q6) and less able to listen easily (Q8).
Participants also rated that in noise, they were less able to share information as desired (Q9) and had to make a higher effort to be understood (Q10).
Note that the normal hearing participant group rated all aspects of conversation success at ceiling values in quiet.
The level of animation affected the experience of being spoken to in a helpful way (Q6) and the ease of listening (Q8). For Q6, in the condition 'transmitted' participants felt more spoken to in a helpful way compared to 'static'.
The main effect of \enquote{animation} on Q8 was not significant in the pairwise comparisons due to the Bonferroni correction.

\bgroup \begin{figure}[phtb]
  \centering \makeatletter
  \includegraphics[width=1\linewidth]{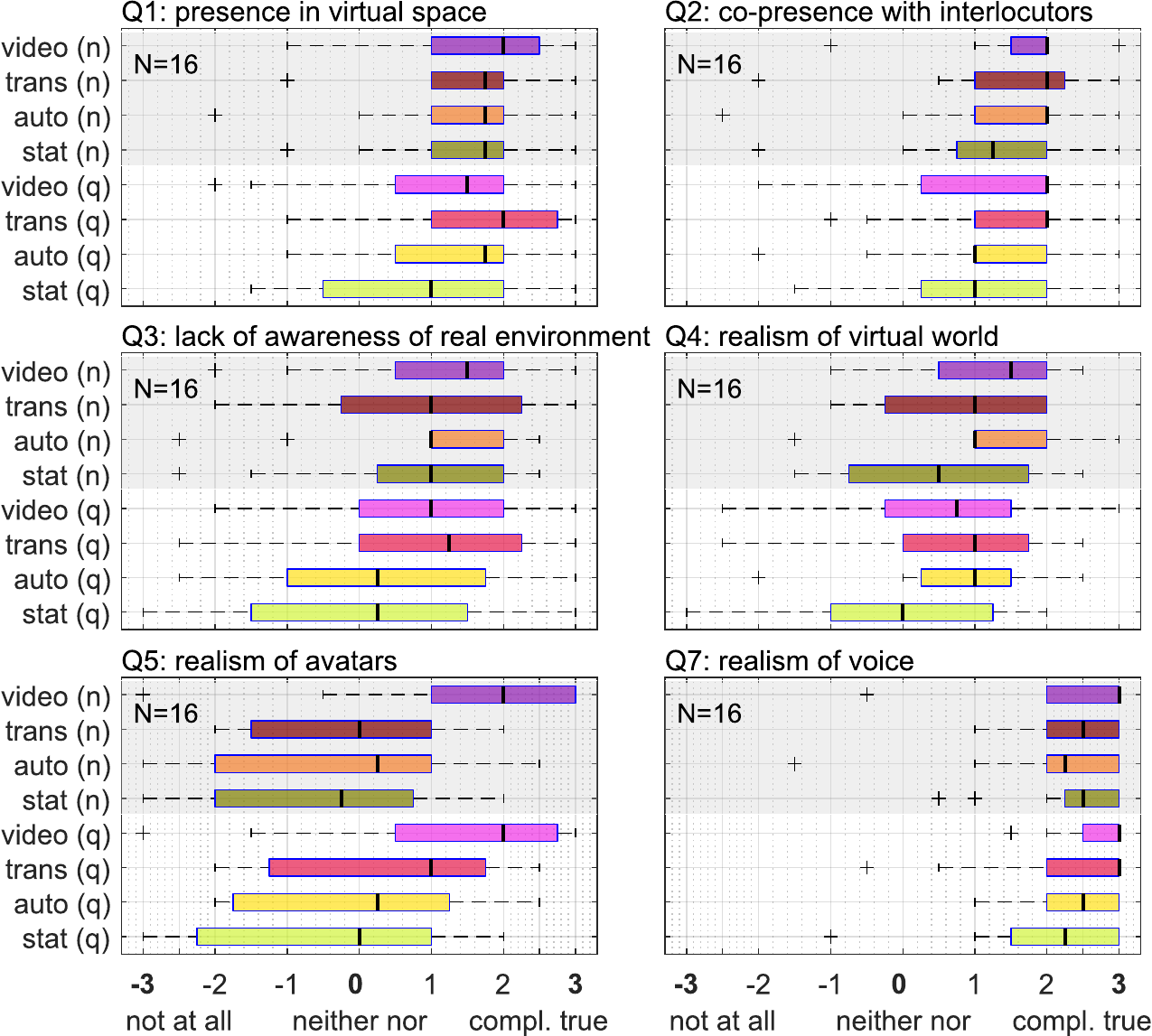}
  \makeatother 
  \caption{\footnotesize Rating values of questionnaire items regarding presence, involvement and realism of the virtual scene. The English labels are parts of the translated original questions in German (cf. Table~\ref{tab:questionnaire}).}
  \label{fig:rating_ipq}
\end{figure} \egroup

\bgroup \begin{figure}[phtb]
  \centering \makeatletter
  \includegraphics[width=1\linewidth]{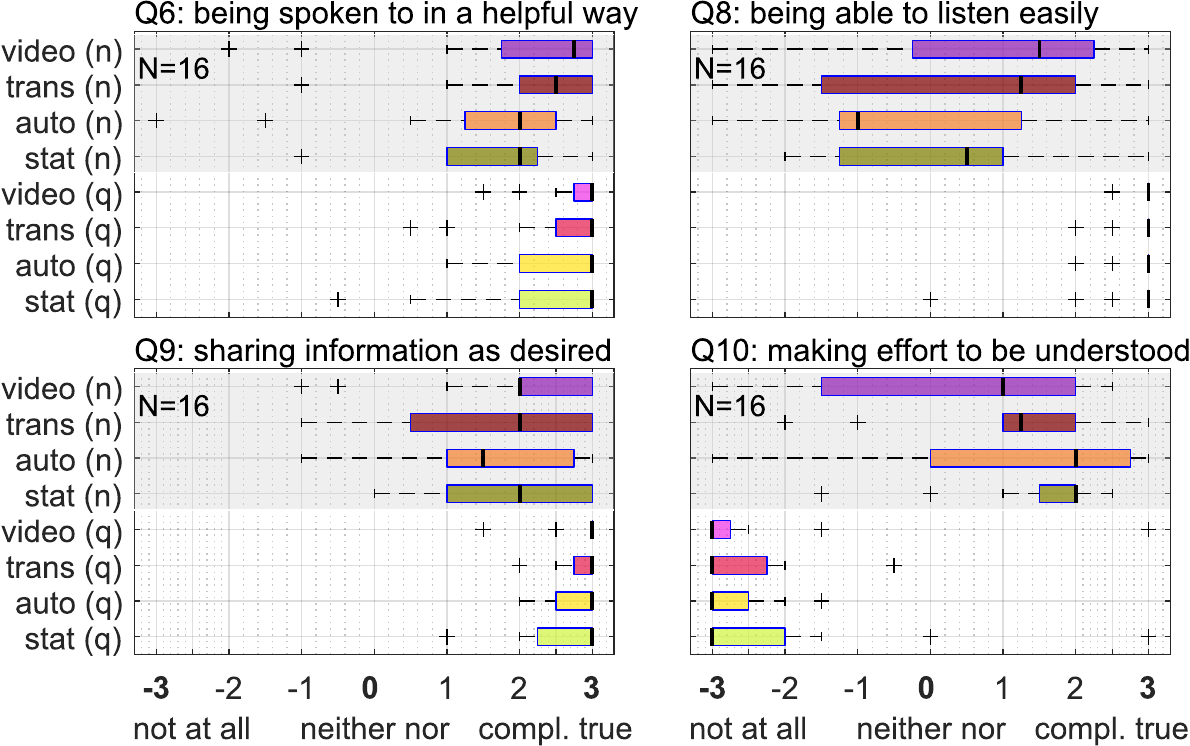}
  \makeatother 
  \caption{\footnotesize Rating values of questionnaire items regarding conversation success. The English labels are parts of the translation of the original questions in German (cf. Table~\ref{tab:questionnaire}).}
  \label{fig:rating_convsucc}
\end{figure} \egroup

\begin{table*}[phtb]
  \footnotesize
  \hskip-3cm\begin{tabular}{|l l|c|c|c|l|r|r|c|r|}
    \hline
    \multicolumn{2}{|l|}{  }                    & \multicolumn{4}{c|}{\bf Animation} & \multicolumn{4}{c|}{\bf Noise}                                             \\
    \multicolumn{2}{|l|}{Variable and Measures}         & $F$        & $p$       & $\eta^2_p$ & \multicolumn{1}{c|}{pairwise comp.}   & \multicolumn{1}{c|}{$F$}    & \multicolumn{1}{c|}{$p$}      & $\eta^2_p$ & \multicolumn{1}{c|}{$\Delta$}    \\
    \hline
    \hline
    \multicolumn{10}{|l|}{{\bf Communication behaviour:}}                                                                                                                         \\
    \hline
    \multicolumn{2}{|l|}{{speech behaviour}}                                                                                                                                     & & & & & & & & \\ 	
              & speech level (participant)             & 3.45       & 0.024     &            & *                         & 692.27 & $<$0.001 & 0.979      & 10.6~dB      \\
\hspace{0.5cm}& {\bf utterance duration (participant)}                & 3.80       & 0.016     & 0.202      & autom | video ($p=0.036$) &\gr{1.52}&\gr{0.237} &           &              \\
              & speech gap duration                     & \gr{1.71}  &\gr{0.179} &            &                           & 6.35    & 0.024     & 0.297     & 0.074~s      \\
              & speech overlap duration                 & \gr{1.11}  &\gr{0.354} &            &                           & 16.28   & 0.001     & 0.520     & $-$0.36~s    \\
    \hline  
    \multicolumn{2}{|l|}{{head movement behaviour}}                                                                                                                                     & & & & & & & & \\                                                                                                               
              & head translation                        & \gr{0.74}  &\gr{0.482} &            &    					   & 8.31   & 0.012    & 0.372      & 3~cm        \\
              & {\bf angular distance to speaker}       & 4.82       & 0.006     & 0.256      & static | video ($p=0.039$) & 14.4   & 0.002    & 0.507      & $-$1.6° \\
              & head orientation range (listening)      & \gr{2.31}  & \gr{0.9}  &            &                            & 17.19  & $<$0.001 & 0.551      & 3.4°         \\
              & {\bf head orientation range (speaking)} & 6.32       & 0.001     & 0.345      & static | video ($p=0.047$) & 6.56   & 0.025    &  0.353     & 3.9°         \\
              &									  &		      &           &             & autom | video ($p=0.005$)  &         &         &            &             \\
    \hline                                                                                                                 
    \hline                                                                                                                 
    \multicolumn{10}{|l|}{{\bf Experience:}}                                                                                                                                       \\
    \hline    
    \multicolumn{2}{|l|}{{ sense of presence}}                                                                                                                                 & & & & & & & & \\
                         & presence in virtual space                    &\gr{1.42}&\gr{0.259}&            &   					    &\gr{2.28}&\gr{0.152}&  &                                                          \\
                         & copresence with interlocutors                  &\gr{0.79} &\gr{0.509}&           &  					    &\gr{2.92}&\gr{0.108}&  &                                                          \\
                         & {\bf lack of awareness of real environment}  & 3.72    & 0.018     & 0.199 & static | video ($p=0.042$)     &\gr{3.04}&\gr{0.102}&  &                                                          \\
                         & realism of virtual world                     & \gr{1.77}&\gr{0.185}& 		 &   						&7.50 & 0.015    &0.333  &0.4~pt.                                                          \\
                         & {\bf realism of avatars}                     & 7.58    & 0.009     & 0.336 & static | video ($p=0.032$)     &\gr{1.43}&\gr{0.251}&  &                                                          \\
                         & 						                     &      &        &       & \gr{static | transm ($p=0.064$)}     &	&	&  &                                                          \\
                         & realism of voice                             &\gr{3.02}&\gr{0.059} &  		  &  					    &\gr{0.05}&\gr{0.826}&  &                                                       \\
    \hline      
    \multicolumn{2}{|l|}{{conversation success}}                                                                                                                                         & & & & & & & & \\                                                                                                           
                         & {\bf being spoken to in a helpful way}       & 4.32 & 0.024 & 0.224 & static | transm ($p<0.001$) & 11.05  & 0.005    &0.424  & $-0.7$~pt.                                                      \\  
                         & being able to listen easily                  & 3.49 & 0.023 & 0.189 & *  					   & 46.60   & $<0.001$ &0.757  & $-2.5$~pt.                                                      \\
                         & sharing information as desired               &\gr{0.99}&\gr{0.390}& &   						   & 16.56  & 0.001    &0.525  & $-1.0$~pt.			 \\
                         & making an effort to be understood		         &\gr{2.95}&\gr{0.068}& &   						   & 120.64 & $<0.001$ &0.889  & 3.5~pt.                                                   \\
    \hline
  \end{tabular}
  \caption{\footnotesize Summary of main effects, with $F$ and $p$ value, effect size $\eta^2_p$, pairwise comparisons, and mean increase in noise compared to quiet $\Delta$. Measures in bold indicate a significant effect of animation level. * indicates a significant main effect, but no significant pairwise difference after Bonferroni correction.}
  \label{tab:effectmeasures}
\end{table*}


\section*{Discussion} 
In this study, we investigated the influence of avatars' head movements on selected measures of speech behaviour and self-motion, and collected feedback via an experience rating questionnaire.  We expected an influence of the level of animation on communication behaviour in background noise conditions, reflecting strategies to compensate for the degraded acoustic information, and to a lesser extent also in quiet conditions. Furthermore, we expected an improved conversation success with increasing level of animation, because head movement cues provide non-verbal information which can be utilised to predict turn takes.

\subsubsection*{Measures of speech behaviour}
The participants, as well as the two confederates, exhibited an increase in speech level of 10.6~dB (C) in the presence of higher background noise, which is a reaction known as the "Lombard effect" \citep{brumm_evolution_2011}. 
The relative increase in speech level was approximately 4.6~dB per 10~dB increase in background noise, which is comparable to values reported for face-to-face conversations \citep{hadley_speech_2019, hadley_conversation_2021}.
The mean speech level difference of 1.8~dB between the confederates \textit{Mar} and \textit{Joa} is assumed to reflect differences in individual speech behaviour, although it is also possible that this discrepancy is caused by a calibration mismatch.
 
The increased level of background noise did not affect utterance duration of the test participants, opposed to \citet{hadley_conversation_2021}, who reported shorter conversational utterances in noise. Only one confederate, \textit{Joa}, showed a decrease in utterance duration in noise.
The observed significant increase in median utterance duration of participants with higher levels of animation can be either caused by overall longer utterances, or fewer short utterances during conversation, such as fewer affirmative utterances during listening.
A trend of fewer utterances by participants with animation level was only present in noise (analysis not shown here).
As expected, no effect on utterance duration was present for the confederates, since they were not exposed to the visual manipulation. 
Alongside with the argumentation of \citet{hadley_conversation_2021}, longer utterances could indicate that, with higher levels of avatar animation, participants shared more complex verbal information and tended to feel less the need to verbally backchannel understanding by affirmative utterances. However it seems contradictory that this facilitated communication was not present in quiet compared to noise. It may be the case that different behavioural patterns were present: for example, long utterances in quiet could contain more complex information, whereas long utterances in noise could be used to keep the conversation going. However this is speculative, as the motivation for longer utterances could not be determined with our data.

Utterance duration values recorded here differed from literature which also investigated free conversations. The median utterance duration of participants over all conditions reported here is 1.08\,s, which is clearly shorter than in \cite{hadley_conversation_2021} with values around 4\,s. The difference in language does not seem to have an influence of that magnitude, since utterances in speech corpora are not necessarily divided by language and have a mean duration of about 2\,s \citep{gonzalez-dominguez_frame-by-frame_2015}. Although our analysis approach as well as the conversation topics are considered to be very similar to \cite{hadley_conversation_2021}, it seems that a different type of utterances is prominent in our analysis. Most detected speech utterances occurred during longer speech sections of another speaker and were relatively short, therefore they can be interpreted as affirmative utterances. Longer sections of speech occurred less often and are not represented in the median value. 

Speech gaps and overlaps between speaker turn takes were not significantly affected by head movement animation. It has to be considered that this measure is based on the interaction of all three interlocutors, of which only one person was exposed to the visual conditions. When all three interlocutors are immersed in the visual environment, this effect can be expected to be larger. Gaps tended to be shorter with higher animation levels in quiet, but not in noise, contrary to expectations. It seems that the non-verbal information did not facilitate the conversation in acoustically adverse conditions. Although head movements have been found to accompany speech prosody and support, for example, syllable identification \citep{munhall_visual_2004}, \cite{saryazdi_gesture_2022} suggested that in noise, a lack of benefit from real-time visual information may be related to occupied cognitive processing in adverse acoustic conditions when the audio-visual setting is complex.

The effect of noise on speech gaps and overlaps was significant: in the noisy conditions, overlaps were shorter and less frequent, and gaps were more pronounced, i.e. longer and more frequent. 
This indicates that the interlocutors took more time for their responses in higher background noise, which has also been found by \citet{sorensen2020timing}. Longer speech gaps are associated with increased cognitive load and processing time due to hindered access to verbal information, and possibly indicates a strategy to facilitate communication in noise \citep{sorensen2020timing}. 

In this study, the magnitudes of gap and overlap duration differ from values reported in the literature. The median duration of speech gaps during conversation found here is 506\,ms, which is approximately 200 to 300\,ms longer than typically reported \citep{levinson_timing_2015, heldner_pauses_2010, sorensen2020timing}, and the median duration of overlaps is 1108\,ms, which is approximately 650 to 900\,ms longer than median values reported \citep{heldner_pauses_2010, levinson_timing_2015, hadley_speech_2019}. Typically reported, and referenced here, are dyadic conversations. However triads were found to show even shorter turn transition times than dyads, which can be related to an increased competition for the conversation floor \citep{holler_competition_2021}. It can be assumed that differences in the duration of speech gaps and overlaps are due to different analysis methods. Level-based detection of speech activity is highly dependent on the time constants in level meters, the level thresholds used to detect the onset of speech, and the possible exclusion of gaps and overlaps above a certain duration, the latter of which was not done here. 

A detrimental effect of the increased transmission delay (measured: $\Delta t=49.8$\,ms) compared to the sound transmission in a physical room ($\Delta t = 340\frac{\mathrm{m}}{\mathrm{s}}/1$\,m $\approx 3$\,ms) cannot be excluded. However this delay lies far below typical video conferencing delays, which range from 120 to 500~ms \citep{tsioutas_audio_2022} and was not expected to have a highly detrimental effect.

\subsubsection*{Measures of head movement behaviour}
The overall undershoot in head angle to the avatars' position of 15.4° to 23.4° is similar to behaviour in in-person settings \citep{hadley_conversation_2021, lu_investigating_2021}.
In an immersive virtual audio-visual environment, the behaviour of orienting the head only partly towards the target direction seems to be maintained, as also reported by \citet{hendrikse_movement_2019}. 
Most participants oriented their heads more closely towards one of the two avatars, see Figure~\ref{fig:movem_indiv}. This avatar represented the experimenter who introduced the participants to the experiment and asked them to complete the questionnaire after each condition. Therefore, an increased attention to this person seems plausible, although this imbalance was not desired in the study design.

In the video condition, participants oriented their heads 2.1° closer to the avatars than in the static condition, possibly because they received more visual information and paid more attention to them. Still, without an analysis of gaze in addition to head movement, this hypothesis cannot be further investigated.
The change in head angle distance suggests that how we visually perceive our interlocutors directly affects our head movement behaviour. This highlights the importance of using study designs that reflect real-life situations when investigating listeners' head movement behaviour in hearing device research \citep{grimm_review_2020}. Despite the modest effect found here, we must be cautious in concluding that there is no significant interaction with directional microphones compared to a static visual setup. This is because directional signal processing also affects our orientation patterns \citep{grimm_review_2020, brimijoin_effect_2014}, which was not tested in this study.

The effect of background noise level on orientation towards avatars was comparable to the effect of video vs. static, with a 1.6° closer mean head angle in noise. 
In contrast, \citet{hadley_conversation_2021} did not find an effect of speech-shaped noise in a level range of 54\,dB to 78\,dB. 
It is likely that the slightly more accurate orientation that was found here was lost in the variance of the interlocutors' position in \citet{hadley_conversation_2021}. Our avatars' heads stayed at the same position, only slightly modulated by the breath animation. This argumentation can also be supported by the results of \citet[Figure~6]{lu_investigating_2021}, who evaluated the head orientation relative to the recorded active talker and found a slightly more accurate orientation of normal-hearing listeners in 70\,dB background noise compared to no noise.
This indicates that target positions have to be tracked if the goal is to precisely evaluate the listener's orientation.

The qualitative analysis of head orientation ranges (5th-95th percentile of yaw angles) revealed individual differences, ranging from 11.2° to 48.3° across all conditions. This great individual variance is illustrated in Figure~\ref{fig:movem_indiv} and should be considered in assessing the benefits of directional microphones. Moreover, different head movement behaviours were observed during listening and speaking, consistent with prior studies \citep{hadley_speech_2019, hartwig_speaking_2021}. During speaking, three-fourths of participants centred their head, likely to engage with both interlocutors.

The head orientation ranges increased with animation level during speaking phases, suggesting a stronger multimodal connection beyond just acoustic communication.  During listening this effect was noticed only as a trend. Furthermore, in noise compared to quiet, both listening and speaking exhibited larger head orientation ranges, indicating a heightened focus on non-verbal cues while seeking additional information.
rbal information is used in the background noise if it is offered.

Participants decreased their distance to the screen in noise by 3.0\,cm. This magnitude is consistent with the literature for seated triadic and diadic face-to-face conversations \citep{hadley_conversation_2021, hadley_speech_2019}, both resulting in a decrease of inter-personal distance of about 3 cm in an increasing noise level. That indicates that also in the virtual environment participants used this universal strategy with the goal to optimise their acoustic situation, or to indicate difficulties.

\subsubsection*{Experience of participants}

The questionnaire revealed that participants felt equally present in the scene in all conditions (Q1, Q2). In contrast, higher animation levels increased the sense of involvement (Q3), supporting the notion that involvement and spatial presence in a virtual environment are distinct concepts \citep{schubert_experience_2001}, and aligning with our expectation that head movements provide additional relevant information in a conversation that require attention.

The rating of avatar realism is plausible, indicating more realistic visual appearance of interlocutors with higher levels of animation and equally realistic voices. It is worth noting that the participants' rating of the visual appearance ranged from the negative to the positive end of the rating scale and increased only slightly from absent to transmitted head movement.
Not expected was a higher experienced realism of the virtual scene (Q4) in noise, which was a significant effect. A possible explanation could be that more resources were needed to participate in the conversation with higher background noise, which were not available to investigate the visual environment. Participants also tended to be less aware of the real environment in noise, supporting this interpretation, however this was not a significant effect. 
The detrimental effect of noise on all four questions regarding conversation success was expected and indicates that the chosen background noise level was adequate to control for conversation success in a normal hearing participant group. The four questions can be grouped into active (Q9, Q10) and passive participation (Q6, Q8). Avatar head movement seemed to have an effect of passive participation, but less on active participation.

\subsubsection{General discussion}
We had expected a general effect of avatar head movement animation level on communication behaviour and conversation success. We saw this reflected in about two thirds of our selected measures. 
Participants oriented their heads more accurately towards their interlocutors, measured by yaw angle distance and overall yaw orientation range during speaking, but did not show a more relaxed head position, i.e. a larger distance between participants' head position and interlocutors. Via the rating questionnaire, participants indicated their conversation partners as more visually realistic and to be more involved in the virtual scene. They reported to be more able to follow the conversation, but indicted to be equally able to share information. For measures of speech behaviour, we cannot confidently conclude a facilitated conversation with higher levels of head movement animation, as speech gaps and overlaps were not affected by animation, and the reason for the increased utterance duration can only be speculated, as discussed above. 

For all selected measures, we found most significant differences between the conditions 'static' and 'video', rarely between 'static' and 'transmitted' or 'automated' and 'video' and never between 'automatic' and 'transmitted'. Still, the ANOVA indicated that the effect sizes $\eta^2_p$ of animation on most behavioural measures and rating questionnaire items were large ($> 0.14$), , according to the guidelines by Cohen 1969, as cited by \cite{richardson_eta_2011}. This indicates that the representation of interlocutors has a considerable influence on the interlocutors' behaviour. Our results also indicate that it may be possible to substitute real head movements with generated movements. Sensible back-channelling of information, e.g. to show attention and to affirm understanding, may be important for a proper involvement of participants into a virtual scene. This behaviour can for example be implemented by mimicking the study participant's behaviour, as suggested by \cite{aburumman_nonverbal_2022}.

We had expected a larger effect size of noise compared to animation. This was found for all selected measures of head movement behaviour and all but one measures of speech behaviour. The effect sizes $\eta^2_p$ of noise on the selected measures, except for utterance duration, were large ($>0.14$). 

Finally, the participants' behaviour and questionnaire ratings also suggest that this conversation in a virtual environment was in part an adequate representation of real face-to-face conversation. The interlocutors raised their speech level in higher background noise similar to face-to-face conversations \citep{hadley_speech_2019, hadley_conversation_2021},
and showed a typical undershoot of head yaw angle to their interlocutors, as found by \citet{lu_investigating_2021} and \citet{hadley_conversation_2021}. Additionally, the rated realism of the virtual world was rated in the upper half of the questionnaire.
Overall, the results indicate a high level of ecological validity in terms of communication behaviour, despite the usage of virtual reality and telepresence.

\subsubsection*{Limitations}
Speech and head movement behaviours are influenced by the number of interlocutors in a group, hearing status and age \citep{lu_investigating_2021, holler_competition_2021, hendrikse_movement_2019}, so the results found here may only be replicated by studies investigating triadic conversations of young, normal-hearing participants.
In this study, only one interlocutor was surrounded by the virtual environment, which may have reduced the effect of visual conditions on interlocutor interaction, measured here by speech gaps and overlaps at speaker turn takes.
As noted above, the magnitudes of the selected speech measures were highly dependent on the level threshold for detected speech activity. Studies using different speech analysis approaches may yield different speech measure values based on the analysis method alone.


\section*{Conclusions}
In this study, we found that the transmission of head movements to avatars affects the behaviour of participants in interactive triadic conversations in virtual reality. In particular, we found small changes in participant's head movement and in the duration of utterances in varying levels of head movement animation. In addition, the sense of presence in the virtual scene and the conversation success were affected. The effect of transmitted head movements was never significantly different from automated head movements, but as a trend an advantage was visible. The largest effects were achieved when using video transmission to represent the remote interlocutors, compared to static avatars. 

It can therefore be concluded that in the context of interactive communication in virtual reality, the representation of interlocutors must include a sufficient amount of head movement, such as nodding or orienting the head towards the active speaker, in order to elicit natural conversational behaviour from the interlocutors. Together with the finding that participants' head movement behaviour differs between speaking and listening, this aspect is relevant to controlled natural listening scenarios for the evaluation of behaviour-based signal processing strategies. Nevertheless, the transmission of head movements alone does not provide sufficient non-verbal communication behaviour and further research is needed into the effect of facial expressions, gestures and posture of the interlocutor on communication behaviour.

\section*{Statements and declarations}

\section*{Statements and declarations}

\subsection*{Acknowledgements}
We thank D. Rothenaicher for her assistance with the data collection.

\subsection*{Declaration of conflicting interest}
A.K., G.G., and V.H. declared no potential conflict of interest with respect to the research, authorship and/or publication of this article.

\subsection*{Funding}
The authors disclosed receipt of the following ﬁnancial support for the research, authorship, and/or publication of this article: This work was supported by the Deutsche Forschungsgemeinschaft (DFG, German Research Foundation) [Project-ID 352015383 – SFB 1330 project B1].

\subsection*{Ethics approval and informed consent statements} 
The study was approved by the Commission for Research Impact Assessment and Ethics of the Carl von Ossietzky University of Oldenburg (approval number EK/2021/068). All participants provided written informed consent before participating in the study.

\subsection*{Data availability statement}
The behavioural and questionnaire rating data, analysis scripts, and supplemental figures have been made available under doi:10.5281/ZENODO.15294859.
The virtual pub environment is published under doi:10.5281/ZENODO.5886987 \citep{grimm_giso_pub_2021}.

\bibliographystyle{plainnat}

\bibliography{references.bib}

\clearpage

\section*{Appendix}

\bgroup

\begin{table}[h]
  \centering
\footnotesize
\begin{tabular}{|c|c|c|}
	\hline
	 & Presented question & English translation \\[0.1em]
    \hline
	Q1& \parbox{3.4cm}{\vspace{0.5em}\enquote{Ich hatte das Gefühl, in dem virtuellen Raum anwesend zu sein.}} & \parbox{3.4cm}{\enquote{I felt present in the virtual space.}}\\[1.5em]
	\hline
	Q2& \parbox{3.4cm}{\vspace{0.5em}\enquote{Ich hatte das Gefühl, mit den Gesprächspartnern in dem virtuellen Raum zu sein.}} & \parbox{3.4cm}{\enquote{I had the feeling of being in the virtual room with the interlocutors.}} \\[1.7em]
	\hline
	Q3& \parbox{3.4cm}{\vspace{0.5em}\enquote{Der reale Labor-Raum war mir nicht mehr bewusst.}} & \parbox{3.4cm}{\vspace{0.5em}\enquote{I was not aware of my real environment anymore.}} \\[1em]
	\hline
	Q4& \parbox{3.4cm}{\vspace{0.5em}\enquote{Die virtuelle Umgebung habe ich wie eine reale Umgebung erlebt.}} & \parbox{3.4cm}{\enquote{The virtual world seemed to be a real world.}} \\[1.5em]
	\hline
	Q5& \parbox{3.4cm}{\vspace{0.5em}\enquote{Wie real erschien Ihnen das Aussehen der Gesprächspartner?}} & \parbox{3.4cm}{\vspace{0.5em}\enquote{How real did the appearance of the interlocutors seem to you?}} \\[1.5em]
	\hline
	Q6& \parbox{3.4cm}{\vspace{0.5em}\enquote{Ich wurde in einer Weise angesprochen, der ich gut folgen konnte.}} & \parbox{3.4cm}{\enquote{I was being spoken to in a helpful way.}} \\[1.5em]
	\hline
	Q7& \parbox{3.4cm}{\vspace{0.5em}\enquote{Wie real erschien Ihnen die Stimme der Gesprächspartner?}} & \parbox{3.4cm}{\vspace{0.5em}\enquote{How real did the voice of the interlocutors seem to you? }} \\[1.5em]
	\hline
	Q8& \parbox{3.4cm}{\vspace{0.5em}\enquote{Es fiel mir leicht, in diesem Gespräch zuzuhören.}} & \parbox{3.4cm}{\enquote{I was able to listen easily.}} \\[1em]
	\hline
	Q9& \parbox{3.4cm}{ \vspace{0.5em}\enquote{Ich konnte Informationen wie gewünscht teilen.}} & \parbox{3.4cm}{\enquote{I was able to share information as desired.}} \\[1em]
	\hline
	Q10& \parbox{3.4cm}{\vspace{0.5em}\enquote{Ich musste mich anstrengen, um besser verstanden zu werden.}} & \parbox{3.4cm}{\enquote{I had to make an effort to be better understood.}} \\[1.5em]
	\hline
\end{tabular}
\caption[Questionniare Items]{\footnotesize Questionniare items that were asked after each condition. The presented questions were in German, consistent with the language of conversation. The English translation is only used for documentation. Answers were given on a 7-point scale (-3 to +3, verbal anchors at ends and centre) with the option to rate in 0.5 steps.}
\label{tab:questionnaire}
\end{table}
\egroup

\end{document}